\documentclass[a4paper]{emulateapj}
\usepackage{epstopdf}
\DeclareGraphicsExtensions{.eps}

\usepackage[backref,breaklinks,colorlinks,citecolor=blue]{hyperref}  
\slugcomment{Submitted to ApJ Letters}
\shorttitle{Clumps versus self-regulated star formation}
\shortauthors{Mayer et al.}

\begin{document}

\author{Lucio Mayer\altaffilmark{1,2}, Valentina Tamburello\altaffilmark{1}, Alessandro Lupi
\altaffilmark{3}, Ben Keller \altaffilmark{4}, James Wadsley  \altaffilmark{4} and Piero Madau
 \altaffilmark{5,3}}
\altaffiltext{1}{Center for Theoretical Astrophysics and Cosmology, Institute for
Computational Science, University of
Zurich, Winterthurerstrasse 190, CH-8057 Z\"{u}rich, Switzerland}
\altaffiltext{2}{Kavli Institute for Theoretical Physics, Kohn Hall, University of California, Santa
Barbara, CA 93106-4030, USA}
\altaffiltext{3}{Institut dAstrophysique de Paris, Sorbonne Universites, UPMC Univ Paris 6 et CNRS, UMR 7095, 98 bis bd Arago, 75014 Paris, 
France}
\altaffiltext{4}{Department of Physics and Astronomy, McMaster University, Hamilton, ON L8S 4M1, Canada; Marianopolis College, 4873 Westmount 
Avenue, Westmount, Montreal, QC H3Y 1X9, Canada}
\altaffiltext{5}{Department of Astronomy and Astrophysics, University of California at Santa Cruz, 1156 High St., Santa Cruz, 
CA, 95064, USA}

\title{Clumpy high-z galaxies as a testbed for feedback-regulated galaxy formation}

\begin{abstract}

We study the dependence of fragmentation in massive gas-rich galaxy disks at $z > 1$ on feedback model and hydrodynamical method,
employing the GASOLINE2 SPH code and the lagrangian mesh-less code
GIZMO in finite mass mode. We compare non-cosmological galaxy disk runs with standard blastwave supernovae (SN)feedback, which
introduces delayed cooling in order to drive winds, and
runs with the new superbubble SN feedback, which produces winds naturally by modelling the detailed
physics of SN-driven bubbles and leads to efficient self-regulation of star formation.
We find that, with blastwave feedback, massive star forming clumps form
in comparable number and with very similar masses in GASOLINE2 and GIZMO. The typical masses are in the
range $10^7-10^8 M_{\odot}$, lower than in most previous works, while giant clumps with masses above
$10^9 M_{\odot}$ are exceedingly rare.
With superbubble feedback, instead, massive bound star forming clumps do not form because galaxies never
undergo a phase of violent disk instability. Only
sporadic, unbound star forming overdensities lasting only a few tens of Myr can arise that are triggered
by perturbations of massive satellite companions. We conclude that there is a severe tension between
explaining massive star forming clumps observed at $z > 1$ primarily as the
result of disk fragmentation driven  by gravitational instability and the  prevailing view
of feedback-regulated galaxy formation. The link between disk stability and star formation efficiency
should thus be regarded as a key testing ground for galaxy formation theory.

\end{abstract}

\section{Introduction}

In the past decade observational evidence has been accumulating that reveals the existence of massive star 
forming clumps
in massive star-forming galaxies at $z > 1$ (\citealt{Cowie1995, Elmegreen2005, Elmegreen2007, Genzel2008, ForsterSchreiber2011}). The masses and sizes estimated from the observations are large, $10^8-10^{10}$~M$_{\odot}$ and up to a kiloparsec, well in excess of those of present-day Giant Molecular Clouds (GMCs).
It has been proposed that such clumps result from the fragmentation of massive gas
disks driven by gravitational instability (\citealt{Noguchi1998, Agertz2009, Bournaud2012}). A large body of numerical simulation work has been devoted to study the
latter scenario. Early studies, which either neglected feedback or adopted inefficient thermal feedback from SNe, 
have been producing systematically giant clumps with masses in excess of $10^8$~M$_{\odot}$ within massive
gas-rich disks  with baryonic masses  $>10^{10}$~M$_{\odot}$ at $z > 2$.  These earlier
simulations, however, suffered from overcooling as they were adopting the notoriously
inefficient thermal feedback (e.g. \citealt{Ceverino2010, Agertz2009}), which 
artificially enhances both disk instability and star formation.
More recently, various works have
revisited high-z disk fragmentation, both with cosmological simulations and simulations of isolated
disk galaxies, finding less 
fragmentation even in massive gas-rich disks, and generally lower clump masses, in the range $10^7-10^8$~M$_{\odot}$ (\citealt{Tamburello2015, Behrendt2015, Moody2014, Mandelker2015, Oklopcic2016}) , which, at high resolution, show significant
substructures (\citealt{Bournaud2014, Behrendt2015}).
 Yet several differences remain among the various published simulations. These are due to a combination of different initial
conditions and/or different properties of simulated galaxies, different star formation and feedback recipes, but
also potentially different underlying hydro  methods and procedures to identify clumps. 
It has also been highlighted how 
some of the most massive observed star forming clumps perhaps do not result from "in-situ" disk fragmentation, rather 
they could be accreted cores of  massive satellites (\citealt{Mandelker2015, Oklopcic2016}).  


Even the recent simulations employ feedback models
that overproduce stellar masses at high redshift by factors of a few relative to abundance
matching constraints, as highlighted by Mandelker et al. 
(2015), suggesting that gas may still cool too efficiently.
Since disk fragmentation depends on net cooling efficiency
\citep{Gammie2001} it follows that fragmentation might still be overestimated.

An exception is \citet{Oklopcic2016}, who uses the FIRE
simulations with multiple stellar feedback processes calibrated to reproduce the stellar-to-halo mass relation from low to high
redshift \citep{Hopkins2014},
but study in depth only one simulated galaxy making it difficult to generalize their results on the clump properties. 
The following question thus  
arises; is widespread disk fragmentation into massive clumps still going to happen in the presence
of a feedback  model capable of truly regulating
dissipation and star formation?

Furthermore, it is unclear how disk fragmentation depends on the treatment of the hydrodynamics.
For protoplanetary disks simulations have shown
that  fragmentation can be poorly captured by cartesian grid-based AMR codes \citep{Mayer2008}, a technique widely used 
in high-z clumpy galaxy studies (e.g. \citealt{Bournaud2011, Ceverino2010}), 
 unless the initial grid
resolution is sufficiently high.
Likewise, SPH, even in its modern incarnations, may suffer from viscous dissipation and
discreteness noise  \citep{Durisen2007}

In this Letter we compare systematically simulations with an "old generation" and a "new generation"
stronger feedback model designed to regulate efficiently star formation, the so called
"superbubble" SN feedback \citep{Keller2015} .
Moreover, we compare results of GASOLINE2 with equivalent simulations carried out with the new lagrangian
mesh-less code GIZMO using the same sub-grid model for star formation and feedback.

\begin{figure*}
\epsscale{0.5}
\plotone{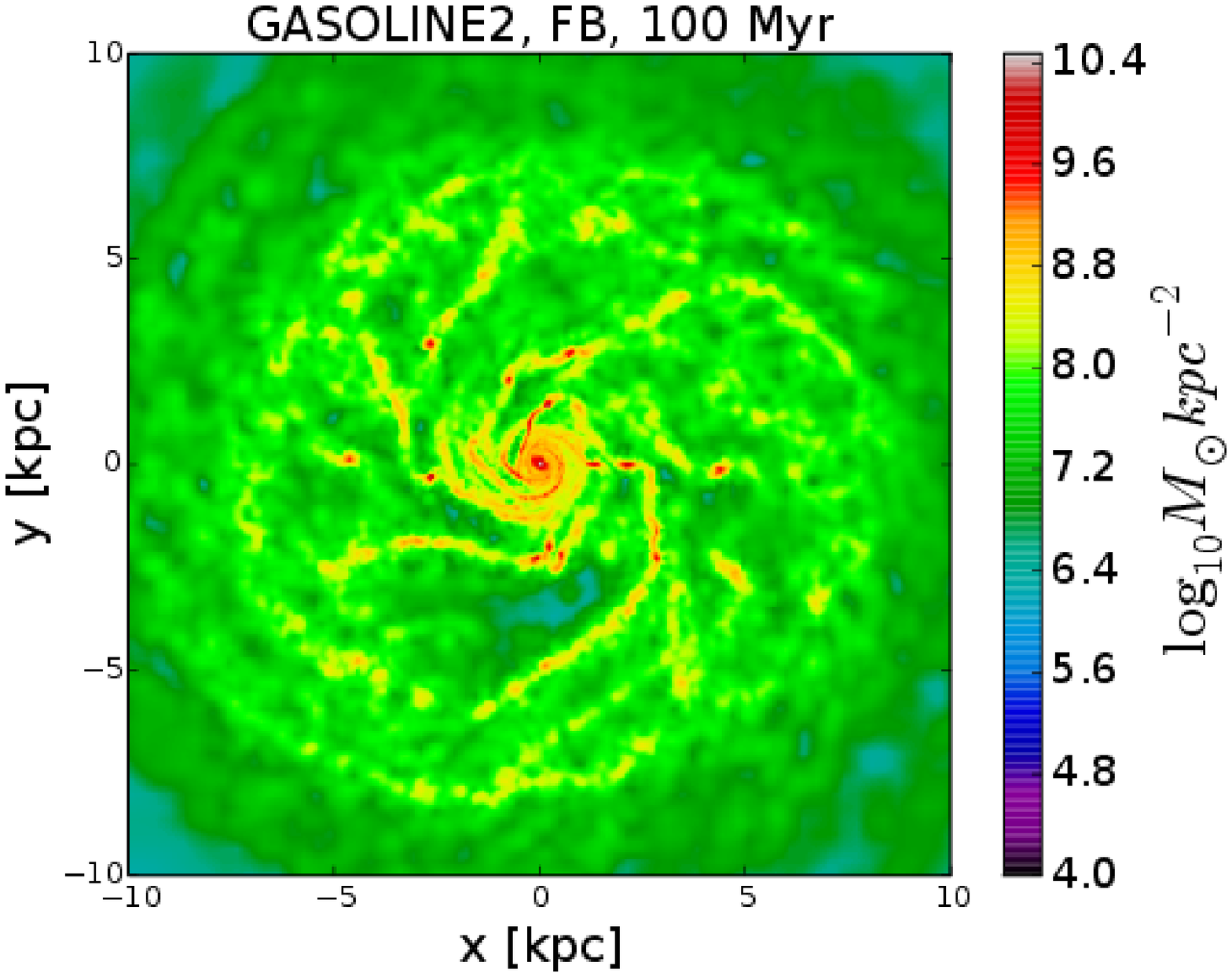}
\plotone{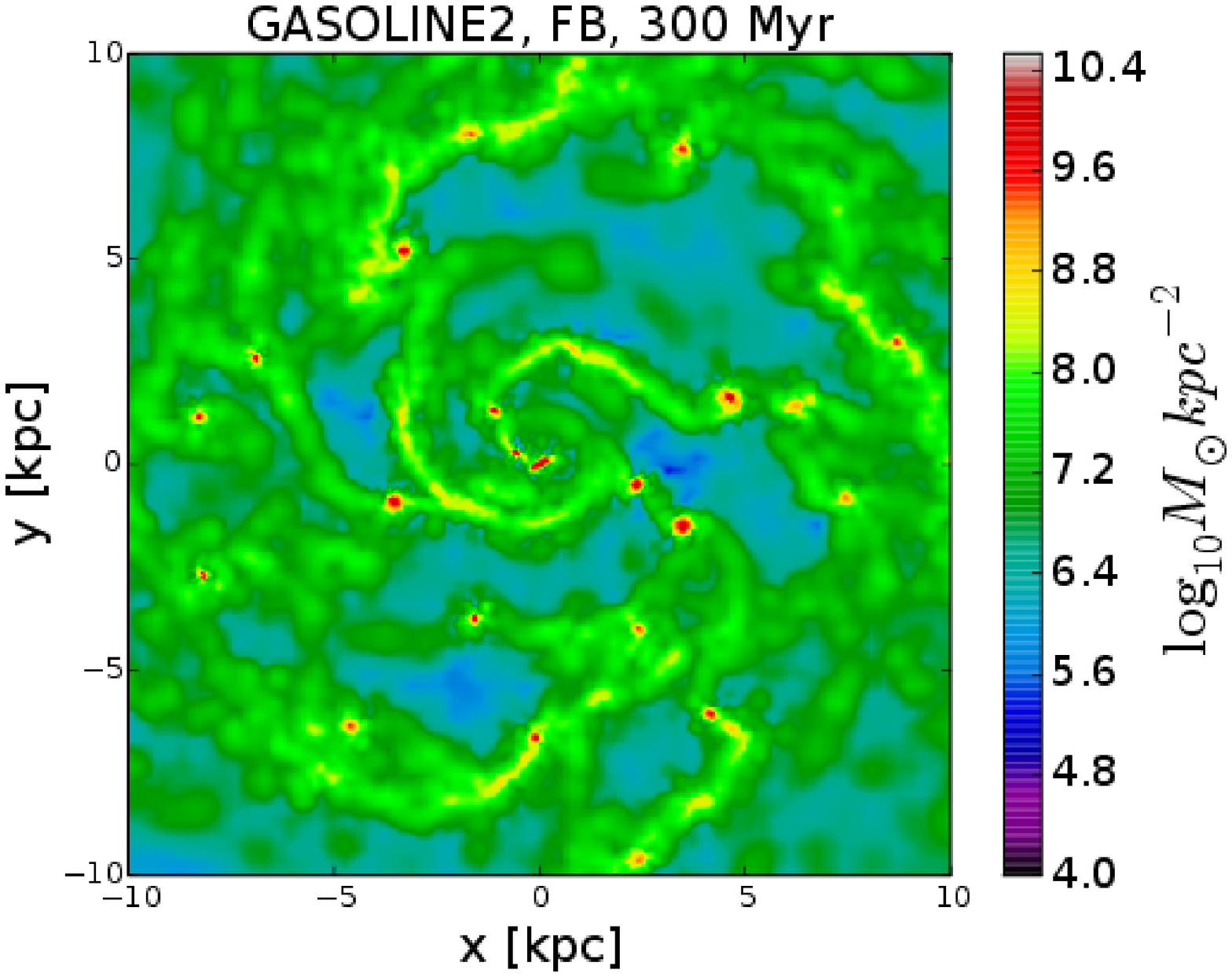}
\plotone{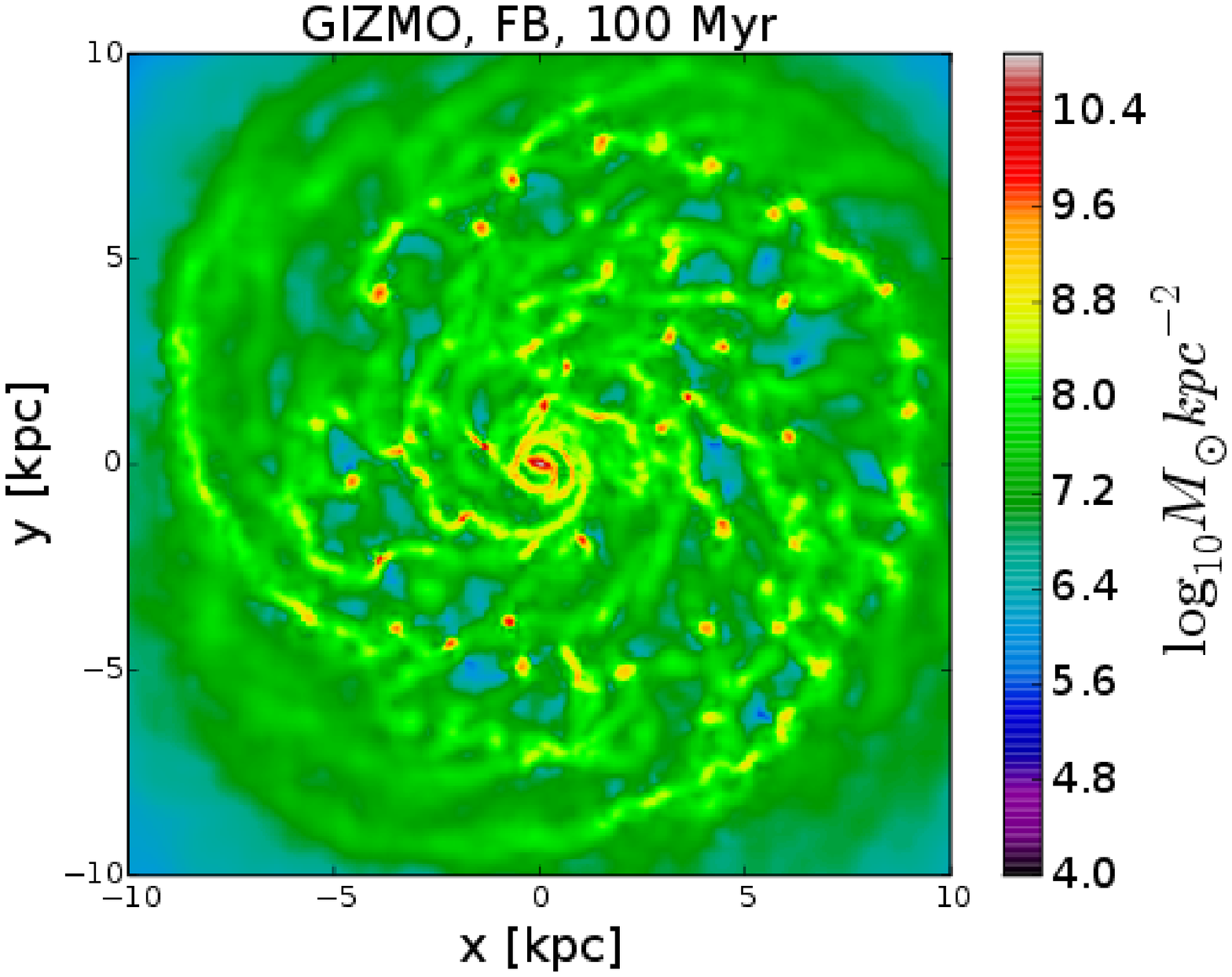}
\plotone{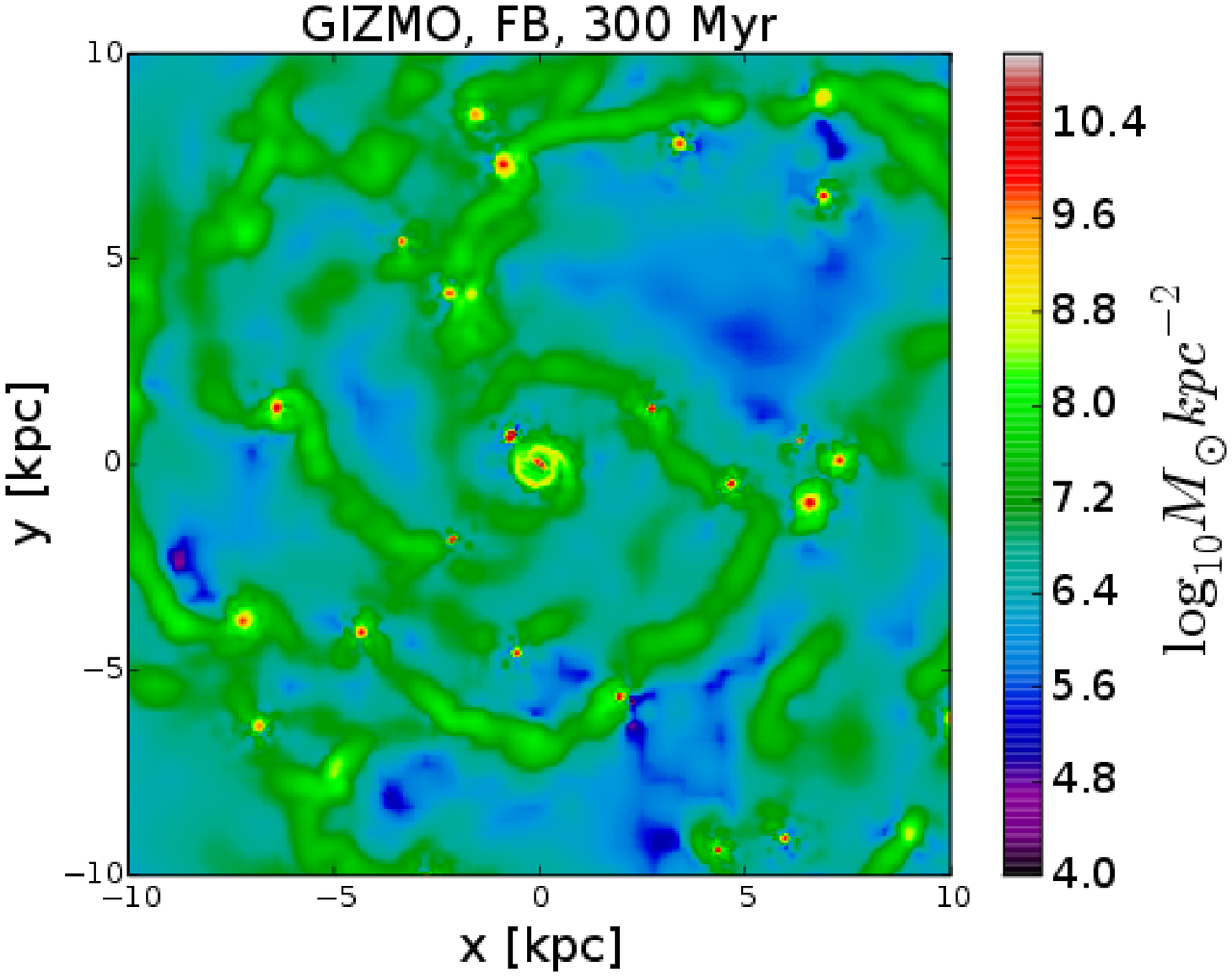}
\caption{Logarithmic color-coded surface density maps of the stellar distribution 
for the GASOLINE2 (top) and GIZMO runs (bottom)  after 100 Myr (left) and 300 Myr (right).
Both runs employ blastwave feedback with the same disk initial condition (the relaxed model 11, see text).}
\label{fig:GizmovsGasoline}
\end{figure*}

\bigskip

\section{Initial Conditions and Simulations}

Since the purpose of our study is to scrutinize the dependence of massive galactic disk fragmentation on hydro technique and sub-grid
feedback prescription we opt to focus on controlled numerical experiments. Therefore we carry out a set of isolated galaxy
simulations. Disk models are constructed as multi-component systems using the standard Hernquist prescription, as 
detailed in \citet{Tamburello2015}. They comprise an exponential disk of stars and gas embedded in an NFW halo, and include adiabatic
contraction.
We adopt two models of very massive high-z disks, designed to yield most favourable conditions for disk fragmentation.
The first model, the default in this work, is the most massive and most gas-rich model adopted in \citet{Tamburello2015}, Model 11. This has a disk of mass $\sim 8 \times 10^{10}$~M$_{\odot}$ with gas fraction of 50\%, and is embedded in halo of mass $2.5 \times 
10^{12}$~M$_{\odot}$. The resulting stellar mass of $\sim 4 \times 10^{10}$~M$_{\odot}$ is high but still within the limits  allowed by
abundance matching constraints. The gas temperature is set initially to $\sim 10^4$ K.
The peak circular velocity, $350$~km/s, is at the very extreme of the line-width of
observed $z > 1$ massive star forming galaxies \citep{Wisnioski2015}. We also construct a second disk model 
that has the same total disk mass and halo mass
but a disk gas fraction of 80\%, at the upper limit of the values inferred from observations at $z > 1$.
The initial conditions (ICs) are relaxed adiabatically to avoid numerical transients which could artificially trigger fragmentation (see \citealt{Tamburello2015}).

\section{Hydro Codes and Sub-Grid Feedback}

We use the GASOLINE2 and GIZMO hydro codes.  GASOLINE2 is an evolution of the GASOLINE SPH
code (Wadsley et al. 2004) which includes an alternative formulation of the hydro force
using geometric density average (GDSPH), a thermal diffusion term in the energy equation and
a Wendland kernel which altogether allow accurate pressure gradient estimates, eliminate
artificial surface tension and allow mixing, thus overcoming traditional limitations
of SPH \citep{Keller2014}. As in \citet{Tamburello2015} we use a pressure floor to
avoid spurious fragmentation. We include non-equilibrium radiative cooling \citep{Shen2010}. 
A companion run with the "vanilla SPH' used in the original GASOLINE code is also
presented for comparison. In addition to adopting standard density-energy hydro force 
formulation and no diffusion the GASOLINE run adopts a spline kernel (Wadsley et al. 2004).
In the GASOLINE2/GASOLINE runs we adopt a fixed gravitational  softening  of 100
pc and a gas mass resolution of $2.26 \times 10^5$~M$_{\odot}$, which
yields hydro resolution as high as 10-20 pc in the high density regions  (a run with ten times
better resolution was presented in \citet{Tamburello2015} yielding similar results).

GIZMO is a mesh-free lagrangian code based on discrete particle tracers that partition the
volume as unstructured cells using a kernel function as in SPH \citep{Hopkins2015}. 
However, unlike SPH codes, these tracers only represent  a
volume partition, sharing an "effective face" with the neighbouring 
ones. The Riemann problem is then solved across these faces, as mesh-based codes, 
allowing excellent shock capturing properties and detection of discontinuities
as in finite volume grid-based codes.
The unstructured cells  are not fixed in space and
time, resulting in a Lagrangian nature of the scheme which allows for intrinsic adaptive 
resolution and an almost exact  conservation of angular momentum (up to gradient errors; 
Hopkins 2015). Gravity is based on the tree algorithm inherited from \textsc{gadget3}, 
descendant of \textsc{gadget2} \citep{Springel2005}, which guarantees high accuracy as well as 
fast computation.
We use a cubic spline kernel function in GIZMO, and the corresponding force softening kernel
function (Hopkins 2015). The gravitational force is softened already at distances smaller than 2.8 
times
the softening length (which is defined as the Plummer equivalent softening) as opposed
to only 2 softening lengths with the spline kernel. Therefore  we rescaled the 
softening
length in GIZMO to 70 pc, in order to have the same spatial resolution for gravity
of GASOLINE/GASOLINE2 which employ a spline or Wedland kernel.
Finally, we implement the same pressure floor as in GASOLINE2, albeit the actual
definition differs owing to the different definition of the kernel size h (which is twice 
that in  GASOLINE2).
In GIZMO we compute radiative cooling using 
\textsc{GRACKLE} \footnote{\url{http://grackle.readthedocs.org}}
a chemistry and cooling library \citep{Bryan2014, Kim2014}. 

In GASOLINE/GASOLINE2 and GIZMO star formation is implemented via a stochastic prescription based on the Schmidt law \citep{Kennicutt1998}, using the 
same criteria as described in \citet{Stinson2006}. Since we are unable to resolve single stars, our stellar particles are assumed to represent 
an entire stellar population with its age, metallicity and a Chabrier IMF \citealt{Chabrier2003}). We allow stars to form in 
regions above a density of 10 atoms/cm$^3$ and below a temperature of 30.000 K, and adopt a star formation efficiency of 0.01, all as in 
\citealt{Tamburello2015}. In all codes we model stellar feedback via type II and type Ia SNe and including also mass loss from low and intermediate 
mass stars. The time of injection for type II SNe and for the mass loss is computed according to the stellar lifetimes as derived by Hurley et al. 
(2000), while the rate of type Ia SNe is modelled according to a distribution of delay times scaling at $t^{-1}$ between 0.1 and 10 Gyr \citep{Kim2014}. For each SN we couple $4\times 10^{50}$ erg to the gas in the form of purely 
thermal energy. In most of our runs we adopt the blastwave sub-grid feedback model \citep{Stinson2006}.
In the latter model we temporarily inhibit 
radiative cooling for gas particles within the SN maximum extension radius $R_{\rm E}$ 
determined by \citet{Chevalier1974} and \citet{McKee1977}. The actual time-scale for the cooling shut-off is 
computed according to the local properties around the stellar particle and amounts to the sum  
of the duration of the Sedov and snowplaugh phases of the ejecta (see \citealt{Stinson2006}).
For all blastwave feedback runs, with both GASOLINE/GASOLINE2 and GIZMO, we switch-off metal-line
cooling because in \citet{Tamburello2015} we found that
it has overall little effect but tends to increase fragmentation at lower, poorly resolved scales.

With GASOLINE2 only we also perform runs with the new superbubble feedback 
implementation, which allows to drive winds without
the need of cooling shut-off \citep{Keller2014}. 
The superbubble feedback model works by depositing thermal energy and ejected
mass from SNe into a brief multi-phase state.  Multiphase particles each have a
pair of separate densities and internal energies, in pressure equilibrium with
each other,  and for which we calculate separately the cooling rate.  Thermal
evaporation within multiphase particles converts them back to a single phase
typically within a few Myr, and is calculated using classical conduction
theory \citep{Cowie1977}.  Resolved hot bubbles then also evaporate
their neighbours using a stochastic evaporation model.  This model allows SN
feedback to deposit energy into a mass determined by physical
evaporative mixing, rather than as a purely numerical parameter dependent
on resolution.
Superbubble allows to achieve efficient self-regulation
of star formation in galaxy formation simulations \citep{Keller2015}. 
In superbubble runs we include also metal-line cooling as its effect on disk stability was not
previously explored separately.

\begin{figure*}
\epsscale{0.8}
\plotone{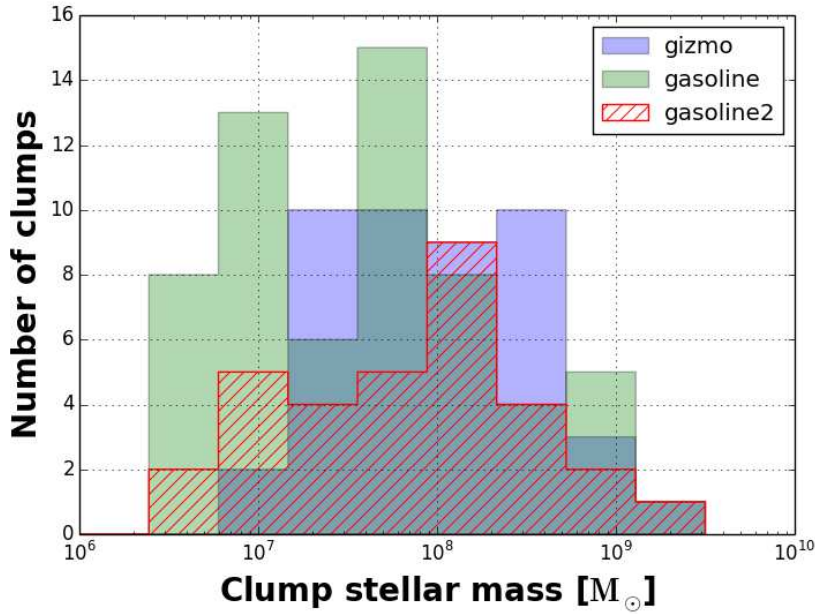}
\caption{Mass function of stellar clumps in GIZMO , GASOLINE and GASOLINE2 massive disk runs with
blastwave feedback, obtained by stacking  different time snapshots (at 200 and 500 Myr)
as in \citealt{Tamburello2015}.}
\label{fig:clumps_mass}
\end{figure*}

\bigskip

\section{Results}

We evolve the disks with cooling, star formation and feedback turned on for at least 500 Myr after the relaxation phase. 
If fragmentation occurs it should happen after only 100-200 Myr, namely
a few local disk orbital times, as expected for gravitational instability \citep{Gammie2001, Durisen2007}. 

Figure \ref{fig:GizmovsGasoline} shows the stellar density maps at different times for runs with GASOLINE2 and GIZMO.
Fragmentation is slightly more vigorous in the GIZMO runs as opposed to the SPH runs, as shown by the 
sharper density contrast between the clumps and the surrounding medium at 300 Myr (Figure \ref{fig:GizmovsGasoline}).
Clumps in GIZMO also exhibit a more elongated
shape. Similar differences have been noted in
the past for sub-stellar clumps in fragmenting protoplanetary disks when comparing SPH and grid-based codes 
(see e.g. \citealt{Durisen2007, Mayer2008}). Overall, clumps span sizes of a few hundred pc in both codes.
When we compare the mass function of clumps identified
using the SKID group finder as gravitationally bound objects, the three simulations yield
similar results (Figure \ref{fig:clumps_mass}).
In particular in the GIZMO and GASOLINE2 runs
the mass function obtained by stacking different time snapshots (see \citealt{Tamburello2015}) 
is flatter and has a less prominent low mass tail relative to the GASOLINE run, but  the characteristic 
mass scale where the distribution peaks, between a few times $10^7$~M$_{\odot}$ and $\sim 10^8$~M$_{\odot}$, 
is comparable in all three runs. Most importantly the  cut-off at $\sim 10^9$~M$_{\odot}$ is similarly sharp 
in all cases. These results confirm the findings of \citet{Tamburello2015} as well as their explanation
of the characteristic initial fragmentation scale based on Toomre-unstable patches in corotation (see
also \citealt{Boley2010}).
The more numerous clumps below $10^6$~M$_{\odot}$ seen in the GASOLINE run are likely the result
of numerical surface tension in standard SPH stifling mixing in regions with high density contrast.
We conclude that, for a given choice of feedback model, results on fragmentation and properties of
massive clumps are relatively insensitive to the hydro method.

\begin{figure*}
\epsscale{0.5}
\plotone{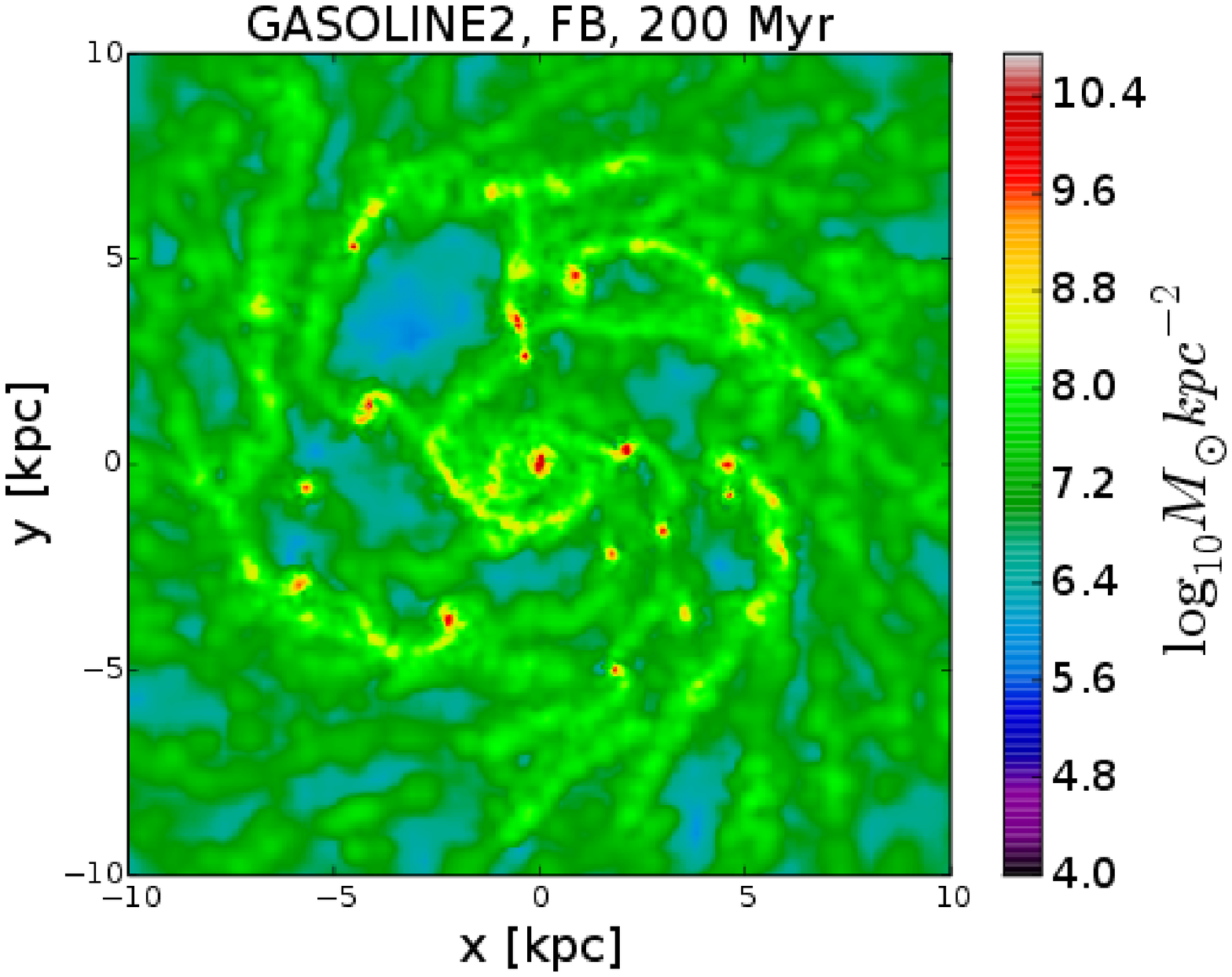}
\plotone{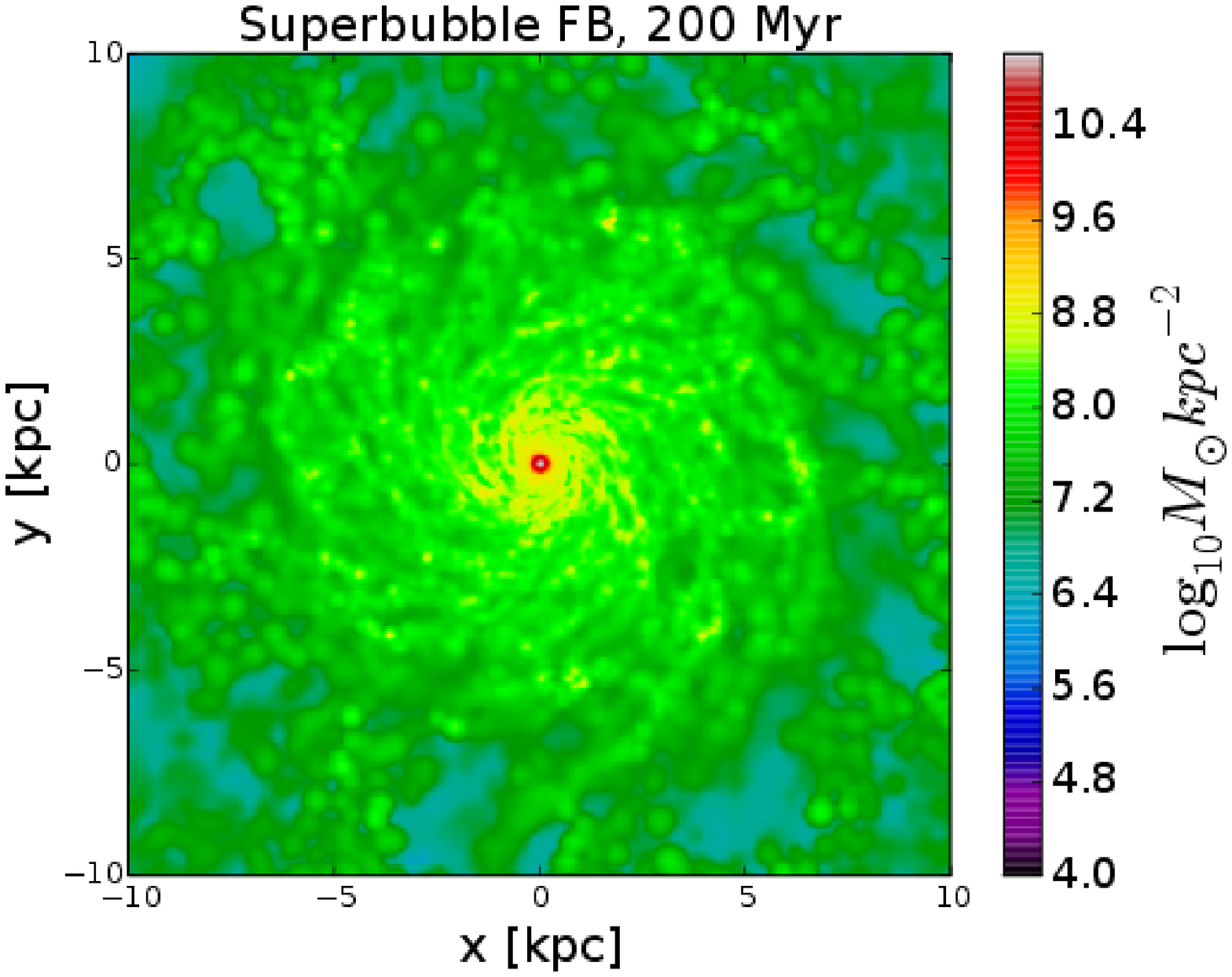}
\plotone{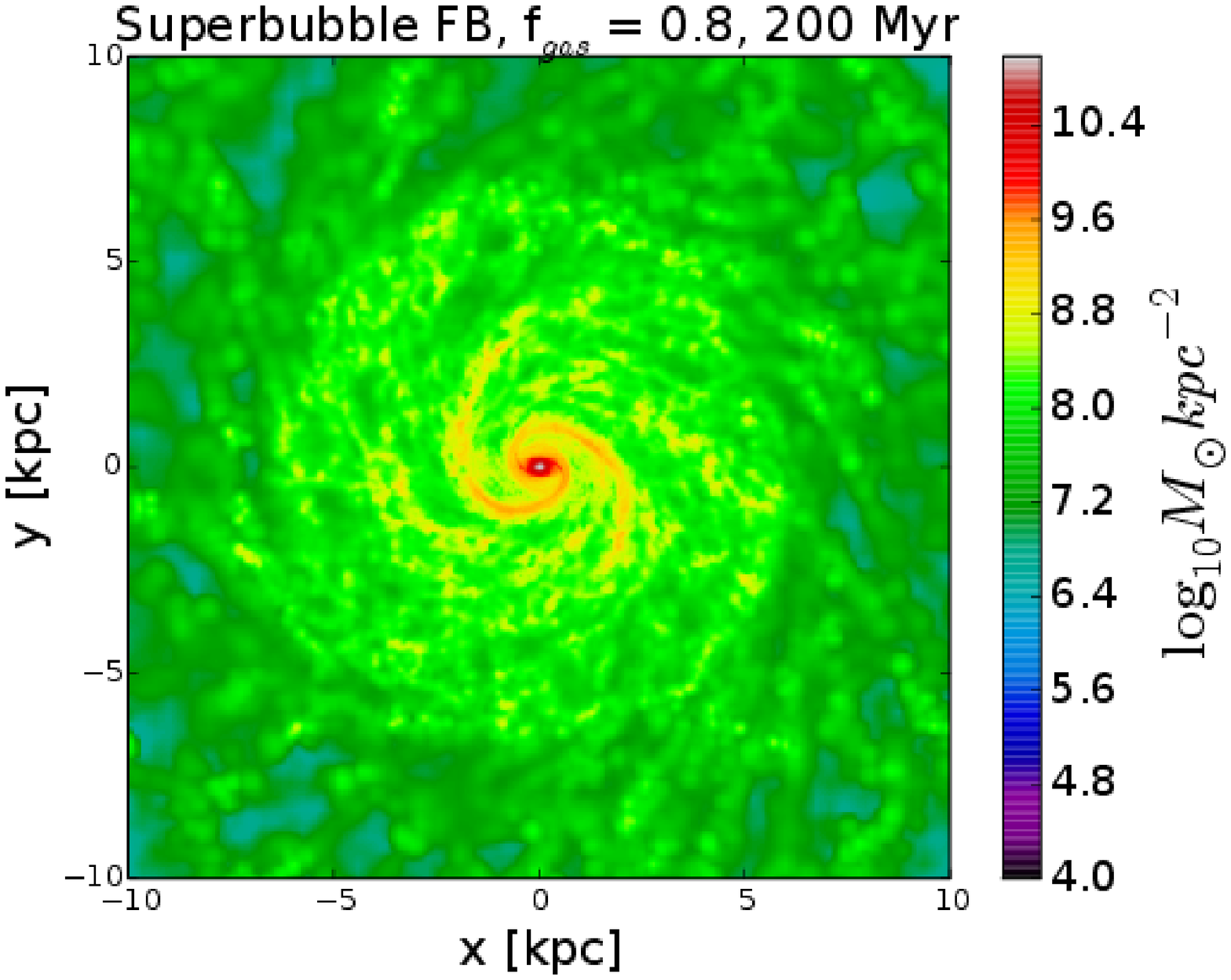}
\caption{Comparison between the logarithmic stellar surface density maps 
for the default galaxy model 11 run with GASOLINE2 using blastwave feedback (left)
and superbubble feedback (middle) at 200 Myr. The model with gas fraction increased
to 80\%, also run with superbubble feedback, is shown in the bottom panel at 200 Myr.}
\label{fig:SBFB}
\end{figure*}

Next we turn to compare GASOLINE2 runs performed with different feedback prescriptions. 
Figure \ref{fig:SBFB} compares the star formation histories of the blastwave and superbubble
feedback runs, highlighting the dramatic ability of superbubble to suppress star formation efficiently. 
There is a factor of several
lower star formation after the initial burst in superbubble. This difference reflects the ability
of superbubble to match naturally the stellar mass to halo mass relation at all redshifts in cosmological
simulations  \citep{Keller2015}.

The suppression
of star formation results from efficient supernovae heating of the gas. Heating acts to stabilize the disk against
self-gravity, thereby leading
to a radically different result compared to blastwave feedback runs (Figure \ref{fig:SBFB}). 
Transient overdensities
do appear only in the first 200 Myr, but disappear in less than an orbit. The lower panel of Figure \ref{fig:SBFB} shows our attempt to verify the
strong suppression of fragmentation by superbubble using the more extreme disk model with 80\% gas mass. This also does
not result in long-lasting clumps. 

We tested that if we reduce the strength of the initial burst by increasing
gradually the star formation efficiency the result does not change, namely the disk still does not give rise to 
giant clumps with superbubble feedback.

\begin{figure*}
\epsscale{0.6}
\plotone{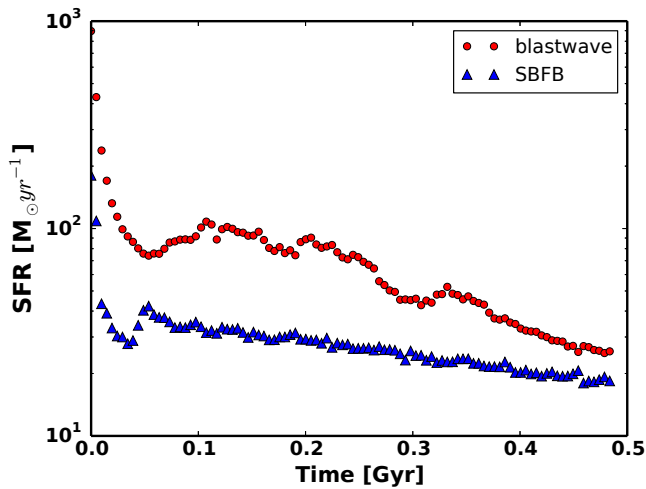}
\caption{Comparison of the star formation histories of model 11 run with GASOLINE2 using blastwave 
feedback (red dots) and 
superbubble feedback (blue triangles). 
The stronger suppression of star formation by superbubble feedback is clearly evident}.
\label{fig:SFR}
\end{figure*}

A limitation of these simulations is of course that they adopt isolated disks, which are neither accreting gas nor can be
perturbed by external galaxies. Both effects are ubiquitous at high-redshift (e.g. \citealt{Agertz2009, Mandelker2015}). Tidal perturbations
might drive disk instability and promote fragmentation in disks that would be stable in isolation. In order to test
this hypothesis we design a new initial
condition in which a massive satellite, with virial mass one fifth of that of the primary galaxy, is launched on a plunging orbit
towards the primary. We use the default disk model 11 for the primary galaxy, and run the simulations with superbubble
feedback only since we want to test how sensitive is the no-fragmentation result to the specific
dynamical configuration. We represent the satellite with a softened massive particle 
having a mass half of its  virial mass (to account for tidal truncation) and a gravitational softening equal to its half mass
radius (radii and masses are scaled from the primary virial mass and virial radius using standard scaling laws
for CDM halos, see eg Tamburello et al. 2015).

\begin{figure*}
\epsscale{0.9}
\plottwo{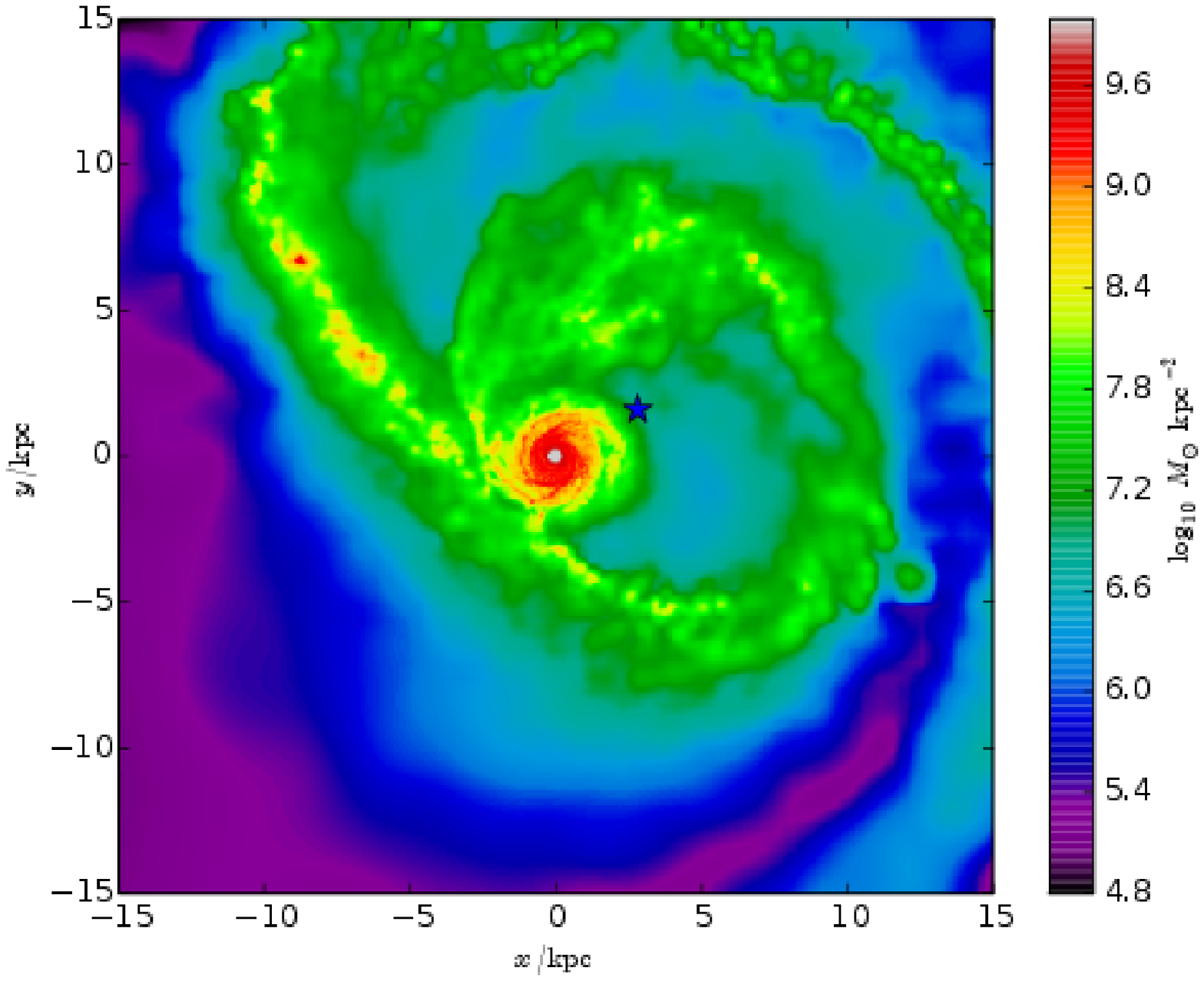}{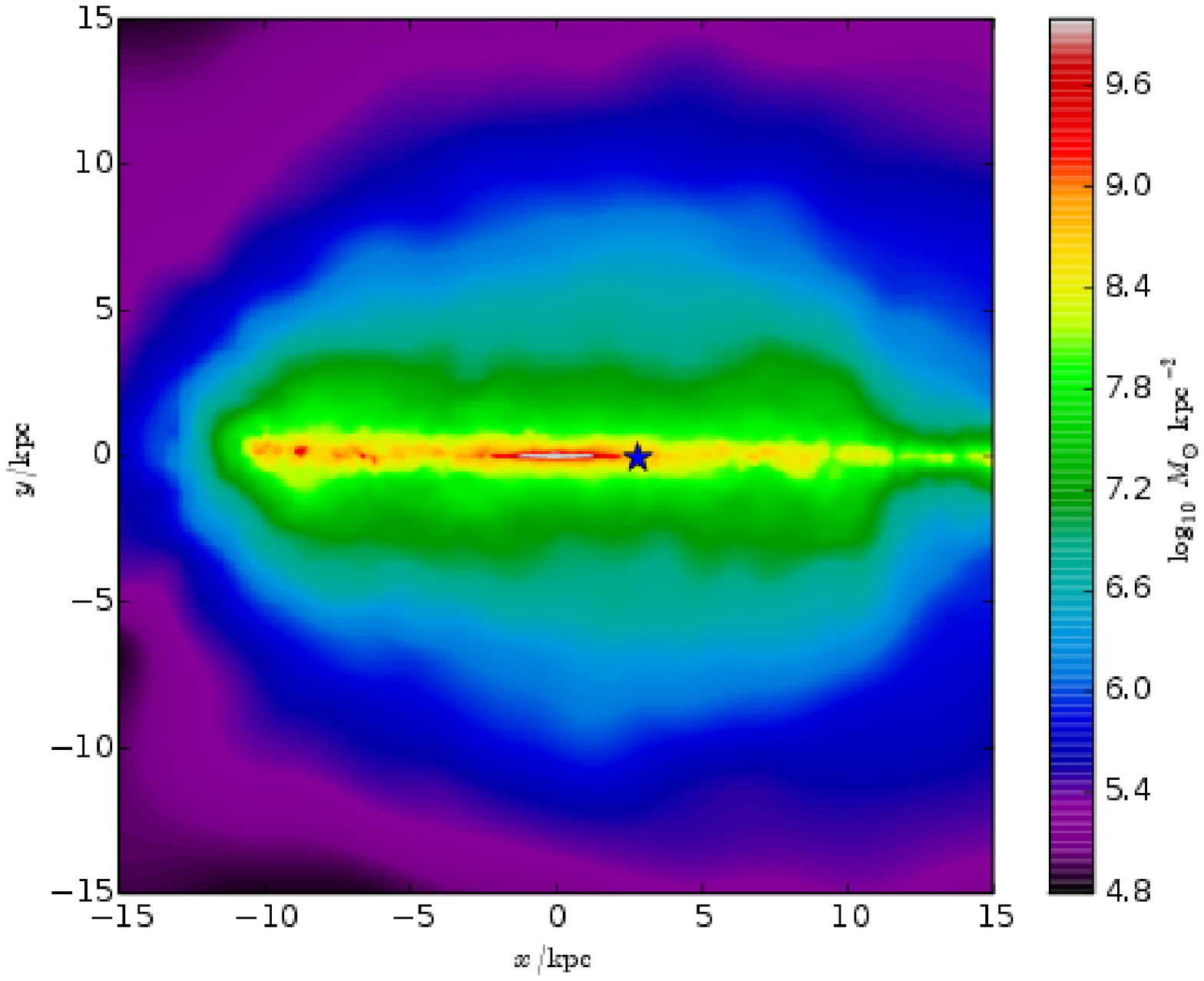}
\plottwo{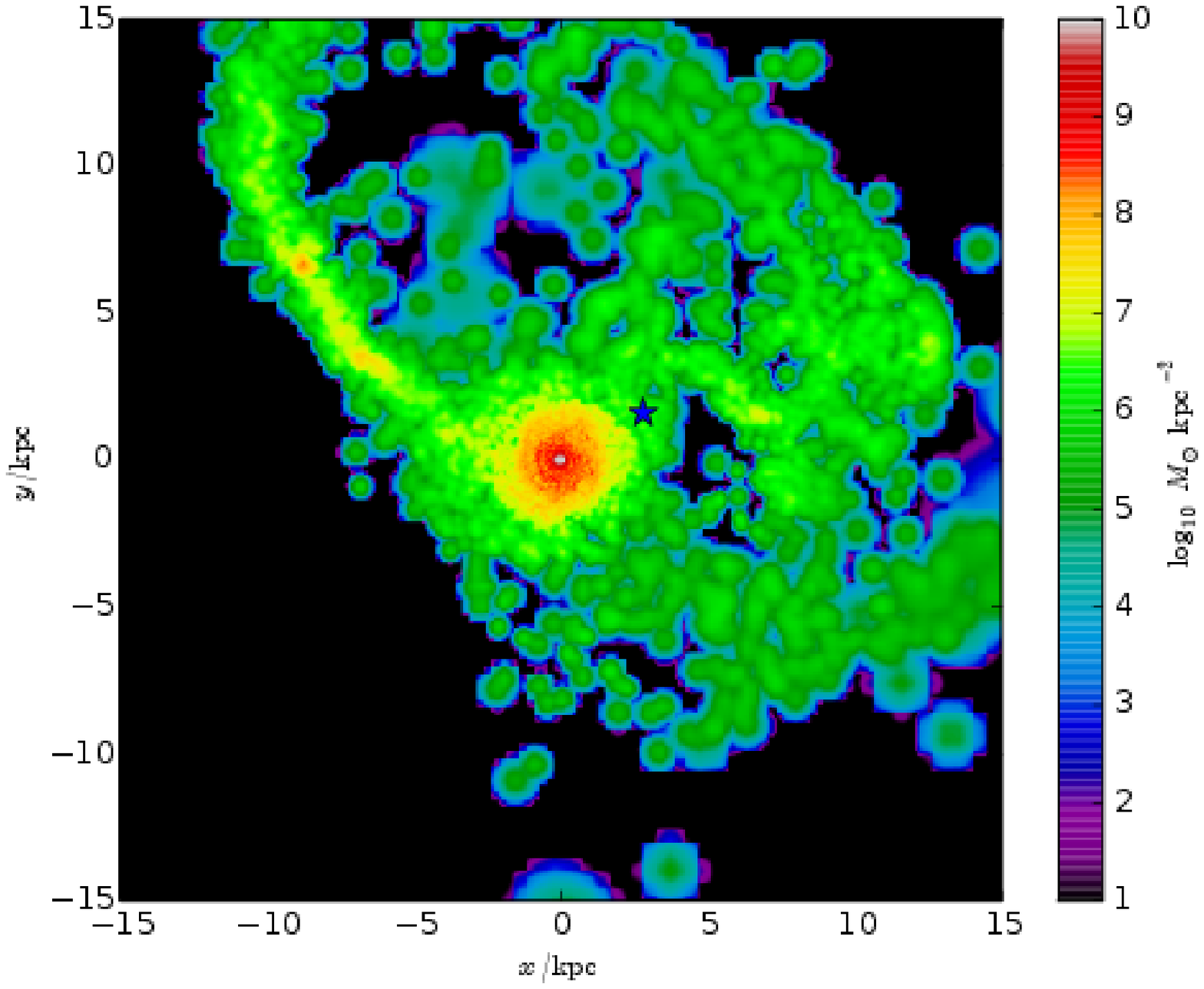}{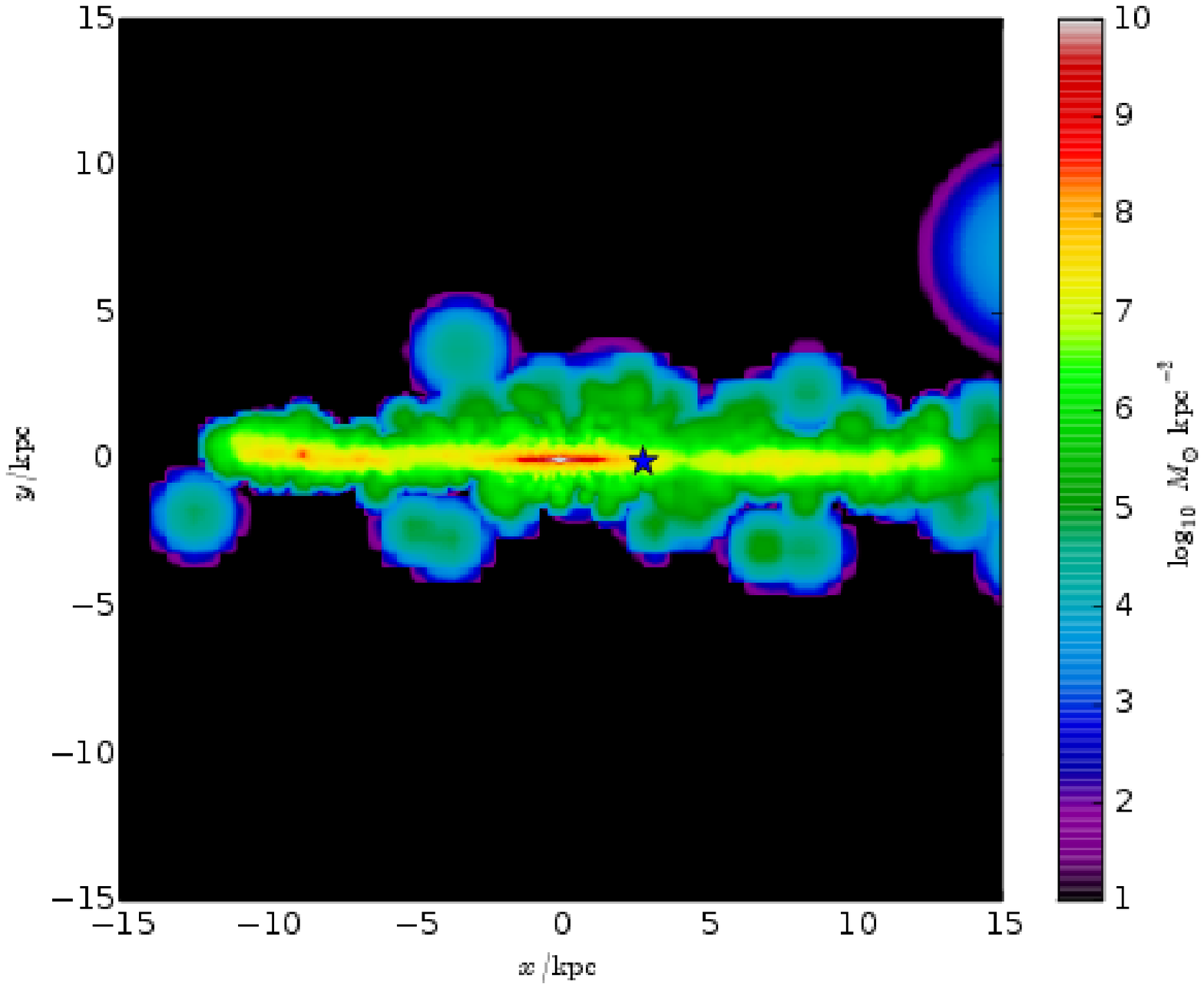}
\caption{Gas (top)  and stellar surface density (bottom, showing only the young stars
forming during the simulation), both face-on and edge-on, of 
the superbubble feedback
GASOLINE2 run in which model 11 is perturbed by a massive companion on a eccentric orbit (see 
text). A snapshot after 190 Myr is shown, about 10 Myr after the formation of the transient clumps, which
can be seen in the upper left corner of the density maps along the tidally induced arm.
The starred point
represents the location of the massive satellite.}
\label{fig:satellite}
\end{figure*}

We run two simulations with the satellite placed on two different orbits, one on an eccentric 
orbit with apocenter-to-pericenter ratio of 5:1 (the pericenter is about 2 kpc), typical
of cosmological substructures, and one on a circular orbit (with a radius of 9 kpc, hence
grazing the outer disk). The simulations 
show that mild fragmentation is triggered in the heavily perturbed outer disk of the primary, although it gives rise 
to only a couple of transient clumps that
last less than 50 Myr. Interestingly, this time scale is comparable to that found by the FIRE cosmological simulations
group  \citep{Oklopcic2016}.
In the eccentric orbit simulation two clumps appear after about 190 Myr, the most significant of which
encompasses $\sim 2 \times 10^8$~M$_{\odot}$ of gas and about $1.7 \times 10^8$~M$_{\odot}$ in stars, and is located 
in the tidal arm forced by the satellite in the primary 
galaxy (Figure \ref{fig:satellite}). It is more prominent in the gas distribution than in the stellar density maps 
(Figure \ref{fig:satellite}). Like this one, all the other triggered clumps are located far 
from the centre of the galaxy, at more than 5 kpc, and are dominated by gas.
Their rapid dissolution is due to the fact that they are loosely bound,
hence easily torn apart by differential rotation as they migrate towards
the inner region of the main galaxy host. This is in good agreement with the results of \citet{Oklopcic2016},
who found nearly all their clumps to be far from virialization. 

\bigskip

\section{Discussion and Summary}

We have presented novel results from a set of simulations employing both different hydrodynamical codes and different
feedback models to scrutinize the popular notion that a significant fraction of observed high-z star forming clumps comes
from the fragmentation of massive star-forming gas disks. We found that fragmentation, including
clump properties, is rather
insensitive to the detail of the hydrodynamical technique adopted, while it is strongly sensitive to 
the feedback
model adopted. In particular, we used a new feedback model that, as some of the other recent feedback models introduced
in the galaxy formation literature, is designed to allow an effective, resolution-independent generation of supernovae
outflows, establishing efficient self-regulation of star formation and allowing to reproduce naturally the
stellar mass-to-halo mass relation from high to low redshift. The result is striking as disk fragmentation is completely
suppressed for isolated disk and only marginally possible, in the form of transient overdensities, in disks perturbed
by massive satellites. The same disks fragment copiously, producing some clumps with masses in 
excess of $10^8$~M$_{\odot}$, when
a less efficient feedback model is adopted, for both
GASOLINE2 and GIZMO. Such feedback model, however, is known to overproduce stellar masses by a factor of 2 or more at 
$z > 0$ for galaxies at the mass scale of the Milky Way and above (e.g. \citealt{Fiacconi2015}).

We argue that our results underscore a new important tension in our understanding of galaxy formation. Violent
gravitational instability of
galactic disks appears to be at odds with the picture of efficient self-regulation of galaxy assembly by
feedback.  Different solutions can be envisioned. First,
it may well be that the outcome we obtain with superbubble is the correct one, 
and that the majority of massive star forming clumps  have an "ex-situ" origin
, for example as cores of massive satellites (in multiple minor mergers, which should be common at high-z
for massive galaxies at the peak of their growth) or from fragmentation of a different nature, such as induced
by thermal instabilities in cold flows feeding the galaxies. {Perturbations by massive satellites would also
be instrumental to trigger some  "in-situ" clump formation, as shown by our superbubble runs with a perturber
and as hinted also by the analysis of \citet{Oklopcic2016}, who showed that the most prominent clumpy
phases correspond to times at which perturbations by satellites occur.
Second, our results might indicate that we are still failing to understand the way feedback works, or at least
how it works at galactic noon, namely at $z \sim 2$, when massive, gas-rich star forming disks are in place and
are at the peak of their growth.  Indeed, the same standard blastwave feedback that overproduces
stellar masses at high z \citep{Fiacconi2015}  does produce
stellar masses that do match observational constraints at low redshift \citep{Guedes2011, Sokolowska2016}}.
Incidentally, work has illustrated how fundamental aspects of feedback, such as
the momentum budget injected in the gas by multiple SNe, has not been correctly predicted so far, suggesting
that the rationale behind the latest generation of feedback models may have to be revised \citep{Gentry2016}. 

In either case star forming massive clumps appear to be a new important testbed  of
our feedback processes in galaxy formation.
Unambiguous evidence of the
fragmentation scenario for high-z star forming clumps will probably come only from direct observations of the
cold gas phase,  since fragmentation starts there in the first place and should leave its imprint on the
dynamics of the gas disk,
for example in the form of large-scale turbulent motions  \citep{Agertz2009b}. 
Hence future observations of the distribution and kinematics of cold neutral hydrogen and molecular gas in $z > 1$ massive 
galaxies may provide the crucial test.

We have considered only two feedback models in this paper. Yet we believe they represent well two
markedly different feedback modes widely employed in the literature. In particular, incarnations of the first one,
blastwave feedback,
have appeared in many of the major hydro codes adopted in galaxy formation, from SPH to AMR
\citep{Kim2014}. Likewise, the second one, superbubble, well represents the
latest generation of feedback models that can achieve naturally self-regulation of star formation at the
desired level, yielding realistic stellar masses as a function of redshift. This is done by 
generating efficient
heating of the gas that reaches temperatures typically $> 10^7$ K as opposed to below that in delayed
cooling models.  More complex feedback models that
achieve the same result but rely on multiple feedback processes, including trapping of infrared radiation
from dust heated by UV to generate significant radiation pressure, as in the FIRE model (Hopkins et al. 2014), 
still generate hot outflows as superbubble. The presumably comparable heating efficiency should lead to a similar 
stabilization of the disk as observed here. Indeed the properties of the clumps in the FIRE simulations, especially
the lack of virialization and short lifetimes, appear very similar to those in our superbubble runs. As already
argued in Tamburello et al. 2015, clumps, if they comprise little mass and are transients,are bound to play a negligible 
role in galaxy evolution, and in particular in the growth of bulges.

\vskip1truecm

L.M. thanks the Kavli Institute for Theoretical Physics  for hospitality, as this work was initiated and completed 
during the Program "The Cold Universe" in Spring 2016. L.M. and V.T. also acknowledge useful discussions with Avishai Dekel,
Manuel Behrendt, Andi Burkert and Dusan Keres.

\end{document}